\documentclass[conference]{IEEEtran}
\IEEEoverridecommandlockouts
\usepackage{cite}
\usepackage{amsmath,amssymb,amsfonts}
\usepackage{algorithmic}
\usepackage{graphicx}
\usepackage{textcomp}
\usepackage{xcolor}
\def\BibTeX{{\rm B\kern-.05em{\sc i\kern-.025em b}\kern-.08em
    T\kern-.1667em\lower.7ex\hbox{E}\kern-.125emX}}

\usepackage{booktabs}
\usepackage{url}
\usepackage{subcaption}

\title{Quantifying How Hateful Communities Radicalize Online Users}
\date{}

\author{\IEEEauthorblockN{Matheus Schmitz}
\IEEEauthorblockA{\textit{Information Sciences Institute} \\
\textit{University of Southern California}\\
Marina del Rey, USA \\
mschmitz@usc.edu}
\and
\IEEEauthorblockN{Goran Muric}
\IEEEauthorblockA{\textit{Information Sciences Institute} \\
\textit{University of Southern California}\\
Marina del Rey, USA \\
gmuric@isi.edu}
\and
\IEEEauthorblockN{Keith Burghardt}
\IEEEauthorblockA{\textit{Information Sciences Institute} \\
\textit{University of Southern California}\\
Marina del Rey, USA \\
keithab@isi.edu}
}

\begin{document}
\maketitle

\begin{abstract}
While online social media offers a way for ignored or stifled voices to be heard, it also allows users a platform to spread hateful speech. Such speech usually originates in fringe communities, yet it can spill over into mainstream channels. In this paper, we measure the impact of joining fringe hateful communities in terms of hate speech propagated to the rest of the social network. We leverage data from Reddit to assess the effect of joining one type of echo chamber: a digital community of like-minded users exhibiting hateful behavior. We measure members' usage of hate speech outside the studied community before and after they become active participants. Using Interrupted Time Series (ITS) analysis as a causal inference method, we gauge the spillover effect, in which hateful language from within a certain community can spread outside that community by using the level of out-of-community hate word usage as a proxy for learned hate. We investigate four different Reddit sub-communities (subreddits) covering three areas of hate speech: racism, misogyny and fat-shaming. In all three cases we find an increase in hate speech outside the originating community, implying that joining such community leads to a spread of hate speech throughout the platform. Moreover, users are found to pick up this new hateful speech for months after initially joining the community. We show that the harmful speech does not remain contained within the community. Our results provide new evidence of the harmful effects of echo chambers and the potential benefit of moderating them to reduce adoption of hateful speech.
\end{abstract}

\section{Introduction}
  
%Social media platforms have had to address important but sometimes conflicting goals: preserving freedom of speech and curbing hate speech and extremism. 
Social media has recently been implicated in a host of societal dysfunctions, such as extremism, polarization, anti-science rhetoric and dissemination of hate. By allowing individuals with particular interests to connect irrespective of physical distances, the web allowed for the creation of countless niches, some of which can be problematic. Reddit (\url{www.reddit.com}) is a social media platform that has become one of the most popular meeting places for such interest-based niche groups, where people create thematic forums called ``subreddits.'' %Although these subreddits can generate new avenues for voices to be heard, they could also be an avenue to incite hate.

One negative externality caused by the ease of congregation in the digital world is that users with racist, misogynist, and similarly uncivil views are able to find others just like them and form communities dedicated to those harmful stances. The formation of such groups can then serve as breeding ground for further deplorable opinions, which, through an echo chamber between community members, may become even more extreme and alienated from reality and good-citizenry, potentially including offline hateful behavior.

This study aims to answer the following research question: \textbf{Does becoming active in a hateful subreddit increase a person's usage of hate speech outside of that subreddit?} Answering that questions will help us understand the impact of hateful communities as originators of hate speech in the platform as a whole.

We create a case-controlled quasi-experimental causal study through Interrupted Time Series (ITS) design and control-treatment user matching. We generate high-precision hate speech lexicons for multiple communities and domains, which are utilized to obtain the measures of hate speech over time. We then find an immediate increase in a user's hate speech as soon as the user becomes active in the hateful subreddit. The causal nature of ITS models implies that becoming active in a hateful community is the factor leading to an increase in hate speech. We observe a repeated pattern of pre-join ramping up in hate speech, and post-join ramping down in hate speech. These findings help us understand whether halting access to the most extreme communities can break the pathway in which users, despite being hateful enough to join a hate community, get even further radicalized after becoming active in such communities. 

% Additionally, we created domain specific lexicons for each of the studied subreddits, which we subsequently leverage to show a difference in vocabulary choice when communicating within the group versus the out-group audience. And lastly, we use the obtained time series data on hate speech to conjecture a qualitative link between the adoption of hate speech and the eventual banning of a subreddit, serving as a reminder of the live interaction dynamics between social media users and moderators.

\section{Related Work}

\textbf{Hate Speech in Social Media.} While it is difficult to devise a uniform definition of hate speech that is both implementable as a computer program while also satisfying critics \cite{silva2016analyzing}, it has been shown that the main targets for hate speech are groups based on race, behavior (i.e. insecure people, sensitive people) and physical appearance \cite{mondal2017measurement,silva2016analyzing}. %, choosing to define it as ``any offense motivated, in whole or in a part, by the offender’s bias against an aspect of a group of people``. Further research by the same group \cite{mondal2017measurement} focuses on stratifying the findings per country and analysing the effects of anonymity. %They identify that the US skews towards more race-based hate speech, Canada sees the highest percentages of behavior-based hate speech, and the UK sees comparatively more sexual orientation-based discrimination. Within the boundaries of the Unites States, race and sexual orientation related hate speech is most predominant in the South, and physical-based hate speech is most predominant in the West. 
Analysis of both directed and generalized hate speech on Twitter has found that directed hate speech is angrier than generalized hate speech, which in turn is angrier than general tweets \cite{elsherief2018hate}. %directed hate speech exhibits the highest clout and the least analytical thinking, while general tweets exhibit the highest authenticity and emotional tone. %Directed hate speech is more informal and social than generalized hate and general tweets. Generalized hate speech emphasizes “they” and not “we”, revealing a tendency to regard in-groups as individuals and out-groups as a mass of undifferentiated people. Both categories of hate speech are more focused on the present than general tweets. Generalized hate has the most death references. Directed hate speech evokes intentional acts, statements and hindering. Generalized hate speech evokes concepts such as People by religion, Killing, Color, People, and Quantity. 
One measurement of hate speech in politically fringe digital communities estimates that one-quarter of all posts contain some form of hate speech, with 13.7\% of posts containing explicit hate speech and 15.5\% containing implicit hate speech \cite{rieger2021assessing}. 
There is a positive association between time spent on Gab (a far-right social media) and hate speech \cite{gallacher2021hate}.
%An investigation on the effects of joining Gab, a social media platform popular with the far-right, found a positive association between the time users spends on the platform and their hate speech \cite{gallacher2021hate}. %While some users join the platform with pre-existing hate stances, others develop these expressions as they get exposed to the hateful opinions of Gab's members. This provides evidence that joining a hate group can increase one's hate speech within that group.
Further exploration of extremist subreddits reveals how those groups combine the the up-vote / down-vote system, Reddit's feed algorithm, governance structure, and anonymity, to surface the extremist content within them and promote extreme discourse against opposing groups, while self-validating their views \cite{gaudette2021upvoting,Massanari2017,gothard2020exploring}.
%Gothard \cite{gothard2020exploring} explores misogynistic communities of Reddit related to Incel (involuntary celibate), many of which have been banned after gaining prominence, and finds the strongest cross-community presence to be within the circle of Incel communities as well as alt-right humor subreddits. The author finds evidence of cross-sharing of content among Incel communities, raising the question of whether individual action against a single community would be able to stop it's members from broadcasting hate speech. 

\textbf{In-Group and Out-Group Behavior on Social Media.} Subreddits are distinctive enough that research has shown it is possible to identify subreddits' posts based on both their style (86.5\% accuracy) and topic (71.1\% accuracy) \cite{tran2016characterizing}. %, the former being a representation of the text with all but the 854 most common words replaced by their POS tags, and the later represented as a 200-category Latent Dirichlet Allocation model. Further, users and discussions with high community endorsement (up-votes) are more likely to match the community's language style. Taken together those findings point towards reveal that (1) each subreddit has its own fingerprint; (2) using the community's language entices greater positive response within the community. 
% Marchal et al. similarly studies the dynamics of in-group vs out-group within Reddit, assessing political polarization in r/Politics, and uncovering 
Moreover, ''the presence of directed political incivility in interactions between dissimilar users tends to discourage further cross-ideological engagement. Conversely, interactions in which ideologically like-minded users engage in derogation of the out-group are less likely to be sustained in the short term'' \cite{marchal2020polarizing}. %While the first finding is in line with conventional expectations of inter-group behavior, the second finding points to a more complex dynamic, where the presence or absence of out-groups, as well as group size differences all modulate the effects that incivility has in inter-group interactions. 
Contrasting the aforementioned findings, in Facebook and Twitter posts about the political out-group were shared more than posts about the in-group, and the number of terms about the out-group increases the odds of a post being shared \cite{rathje2021out}. %, such that out-group language is the strongest predictor of sharing, surpassing negative affect language and moral-emotion language, both established predictors of social media engagement. Further, on Facebook, where users can give specific emotion reactions, references to the out-group was a strong predictor of ``angry`` reactions and references to the in-group was a strong predictor of ``love`` reactions. The authors conclude that ``out-group language is the strongest predictor of social media engagement across all relevant predictors measured, suggesting that social media may be creating perverse incentives for content expressing out-group animosity``.
We note group dynamics are not unique to social media, as a singular analog example, research has shown in-group favoritism and out-group derogation among Europeans and Maoris, but for different reasons: ethnic identity for Maoris and social dominance orientation for Europeans \cite{hamley2020ingroup}.

\textbf{Hate Speech Identification and Prediction.} Another direction of research seeks to leverage machine learning to classify and predict hate speech at scale. Successful approaches have leveraged transfer learning \cite{almerekhi2020investigating} and moderation-minded work \cite{habib2021proactive}.
%to train machine learning models capable of detecting hate speech with a crowd-sourced dataset of toxic posts . %Two sets of models are proposed, one for classifying the toxicity of a post, and the second for predicting user's toxic posting behavior. 
%Such model development 
%This work was followed by more moderation-minded work, assessing the feasibility of proactively identifying problematic communities on Reddit. The authors conclude that deploying tools for such purposes is a promising strategy as it can reduce manual labor associated with community management, provide a scientific rationale for the moderation decisions while also facilitating quicker and softer interventions \cite{habib2021proactive}.
Multiple researches have built datasets for hate speech identification \cite{mathew2020hatexplain,qian2019benchmark,mollas2020ethos,de2018hate}, in addition to building a dataset, De Gilbert et al. \cite{de2018hate} used Pointwise Mutual Information (PMI) comparison between hateful and non-hateful sentences, finding the most hateful words are derogatory or refer to targeted groups of hate speech. %On the other hand, the least hateful words are neutral in this regard and belong to the semantic fields of Internet, or temporal expressions, among others. This shows that the vocabulary is discernible by category [...]``. 

\textbf{Moderation of Hate Speech.} %In the realm of evaluating moderation approaches for web platforms, Chandrasekharan et al. \cite{Chandrasekharan2017} conclude that Reddit's 2015 ban of high profile hateful subreddits resulted in vastly increased churn amongst the members of those communities, while the staying members did not cause an increase in hate speech in the communities they migrated to. 
Various moderation approaches could help reduce the amount of hate speech, including the banning of high-profile communities on Reddit \cite{Chandrasekharan2017}, as well as employing somewhat softer alternatives to bans, such as quarantines \cite{chandrasekharan2020quarantined,copland2020reddit}. Quarantines also led a portion of the communities' members to leave Reddit for more lightly regulated platforms, which merely forwards the issue of handling hate speech to another platform \cite{copland2020reddit}. Those findings corroborate with previous study showing that subreddits influence one another, and thus intervention in one subreddit can be expected to have additional effects on other subreddits, especially those with a large shared user-base \cite{Zannettou2017}. After a subreddit's ban, top users become less active while most users reduced their community-derived language without a reduction in activity. Moreover explicitly racist communities had a stronger reduction in activity from bans than dark humor communities \cite{trujillo2021echo}. Previous research also shows that hateful users differ from regular users, e.g., in their vocabulary as well as engagement \cite{Chatzakou2017}, hence to the extend top users of hateful subreddits are the most hateful users, this can at least partly explain the diverging fates of the top users in relation to regular members.
%An examination of softer alternatives to bans on social media under two case studies of Reddit's ``quarantine`` policy, which prevents direct access to and promotion of controversial communities, finds quarantining reduced the influx of new members to the communities and the spread of hateful content from quarantined subreddits into the rest of Reddit, but did not reduce the levels of hate speech for existing users \cite{chandrasekharan2020quarantined,copland2020reddit}. 

% \textbf{Digital Migration of Hateful Communities.} Both as result of moderation initiatives, as well as due to a myriad other changing factors, it has been observed that certain communities can migrate, sometimes with the intent of avoiding moderation, and sometimes for other reasons.  Davies et al. \cite{davies2021multi} look at Reddit user migration in the context of the COVID-19 pandemic and find that at the user level there is a strong overlap of user communities with a notable interplay of controversies, humor and politics. %, and at the subreddit (community) scale there is increasing politicisation of the pandemic, which is accompanied a rise in conspiracy theories. 
% Newell et al. \cite{newell2016user} studied user migration in Reddit, finding that a factor which helps Reddit maintain users is its long tail of niche content, which can only be lively due to the size Reddit's user base. This indicates that whether a hateful or fringe community can be exterminated by a Reddit ban is related to the community's prominence and its ability to attract new members when residing in less popular platforms.

\section{Methods}\label{sec:methods}
\subsection{Identifying Communities of Interest}
To select the communities of interest we identify subreddits that were known to be a breeding ground for hateful speech. We used previously published research on hateful communities on Reddit and selected the following subreddits: r/GreatApes, r/CoonTown, r/Incels and r/fatpeoplehate as they appear to be the most notorious ones~\cite{Chandrasekharan2017,Ko2021,gothard2020exploring}. All the selected subreddits were already banned when we started our analysis. 

\subsection{Hate Speech Lexicons}
We employ a pragmatic definition of hate speech, with the intent of making hate speech detection a problem that can be defined in algorithms and computational tasks. We broadly define hate speech as usage of words that direct hate and derogation towards a specific group of people based on their gender, appearance, or some other characteristic.

We create a corpus of candidate hate words by employing Sparse Additive Generative Models of Text (SAGE) \cite{Eisenstein2011} to compare all posts within each studied subreddit versus a global sample corpus for all of Reddit. The global Reddit corpus contains a sampling of content published through the entire platform to serve as a baseline of speech over all of Reddit. It consists of 10GB of randomly crawled posts, which were obtained directly through the Reddit API's ``fetch random`` functionality, and includes both submissions and comments. The subreddit specific corpus contains all posts made inside the studied subreddit. We remove stopwords from both corpora. Many hate words are slangs which are not well handled by stemming or lemmatizing, hence we do not apply any such techniques. SAGE then calculates the most distinctive words for the target corpus, which provides a list of unigrams mostly characterizing of each subreddit.

We select the top 300 candidate words and use a 3-rater system, where the authors of this paper served as the raters. Each rater independently classified each candidate word as ``0 - Not a hate word'', ``1 - Might be a hate word'', and ``2 - Always a hate word''. The purpose of this approach is to reflect that certain speech cannot be certainly classified as hateful in the absence of context \cite{cervone2021language,paasch2021insult}. We then sum up the ratings and select only words which score 4 or above to form each subreddits' hate speech lexicon, ensuring words are likely to be hate words. %These lexicons allows us to calculate the percentage hate speech for each user on each day by diving the hate speech count by the total word count for that user-day. 
These lexicons allows us to calculate the level of hate speech - the number of hate words divided by the number of all words used by a user. The level of hate speech is calculated for each user for each day the user has been active on the platform. 
%For each studies subreddit, Figure \ref{fig:WF} displays a pie chart summarizing the most frequently written hate words in the in-group and out-group contexts, considering posts after the users became active members of the hateful subreddit. 
This approach to lexicon construction has the upside of allowing us to capture hate speech that is unusual and in-group specific, thus oftentimes being encrypted to the general population and hence masked from generalized lexicons \cite{Does2022,Gerrard2018}.

\subsection{Data Collection}
\textbf{Treatment Data}. To obtain the post history for all active members of each hateful subreddit we used the PushShift API \cite{Baumgartner2020} to crawl all available submissions and comments made on each subreddit. %It is no longer possible to obtain subreddit data from the official Reddit API as all the studied subreddits have been banned. 
Once all available data was obtained, we then parse through the data and obtain the usernames of all users who posted on each subreddit, those will be the treatment samples for our analysis. Having the usernames we once again leverage PushShift to now crawl the entire posting history for those users. We seek to remove automated posting bots from our sample by searching usernames for the following keywords: ''bot'', ''auto'', ''transcriber'', ''gif'', ''link'', ''twitter'', which were obtained by manual inspection of usernames in the largest files crawled. We manually analyze the usernames matching those keywords by inspecting a sample of their Reddit posts and subsequently remove accounts identified as bots. For each subreddit this resulted in the removal of few accounts ($< 1\%$), yet always at least the top 5 accounts by activity levels were removed, as they were consistently bots. The resulting dataset contain 3140 members of r/GreatApes, 8927 members of r/CoonTown, 17859 members of r/Incels, and 21465 members of r/fatpeoplehate.

\textbf{Control Data}. To increase the probability of choosing ''similar'' users in the control sample, the control candidates are selected from users who are likely to share the topics of interest with the treatment users. Therefore, we select control candidates from the users who are the members of the subreddits frequently visited by the treatment candidates as well, but not members of the studied hateful subreddit itself. We refer to them as the ''control subreddits''\footnote{Not to be confused with the control users. ''control subreddits'' are used as an aid to identify control users that are likely to match with the treatment users.}. To identify the ''control subreddits'' we find subreddits that have the high proportion of already identified treatment users. That way we select the top 30 ''control subreddits''. We then gather all users active on those subreddits as the set of candidate control users.

\textbf{Banned Subreddits}. With the goal of producing additional analyses, we obtain a non-exhaustive list of banned subreddits. Reddit does not divulge its bans, and hence all information comes from user generated content and self-reports. The gathering process was entirely manual and consisted of browsing Reddit itself for posts compiling subreddit bans. Given the nature of the data gathering process, prominent bans and bans of large subreddits are more likely to be featured in our set. In total we obtained 3515 subreddits reported to be banned. 

\subsection{Matching}
Using Mahalanobis distance-based matching \cite{Stuart2010}, we find a 1:1 pairing between treatment users and control candidates, such that the control candidates which are most similar to treatment users are selected. We limit our set of treatment users to 15k users, randomly chosen amongst all active subreddit members, and limit the control candidates to a 5:1 ratio on the actual number of treatment users obtained.

We considered the following set of features: \textit{account creation date}, \textit{Reddit karma (sum of all up-votes minus all down-votes)}, \textit{total number of submissions}, \textit{total number of comments}, and the \textit{count of posts made in each of the top 50 subreddits} (those with the highest ratio of treatment members, as per the list generated when defining the subreddits from which to sample control candidates).

%, which measures distance across a feature space whose dimensionality matches that of the feature set yet with axes shifted onto being the matrix's eigenvectors such that variance is maximized within the axes (as in Principal Component Analysis), and thus covariance is minimized. That feature space is then normalized to the variance of each axis by scaling each axis by its principal eigenvalue, so that distances across all axis have the same weight in the distance calculations \cite{Stuart2010}.

To ensure that the matching procedure does not influence the later analyses, matching only considers users' behavior (i.e. features) prior to the moment they join the hateful subreddit, as joining is akin to beginning treatment \cite{Ham2022}. 
% Given the treatment start date (defined as the date of joining the hateful subreddit) is unique for each user, we create a 3-dimensional data structures for the treatment users and another for the control candidates. On the first axis we have each of the users, represented by their usernames. On the second axis we have each of the features. On the third axis we have months covering the entire range of join dates for treatment users. On each location in the cube we have the cumulative values of the features for a user up to a given month. This structure allows for the following procedure: 
The matching algorithm uses the following procedure: (1) select a treatment user, (2) check in which month that user became active on the hateful subreddit, (3) consider that user's features and the control candidates' features on the month prior to the activation month, (4) find the most similar control candidate via Mahalanobis distance matching, (5) store a triplet of (treatment, control, distance), (6) move to next treatment user.

% There is a possibility that more than one treatment user gets matched to the same control candidate, hence once all users were matched, we check for control candidates who had more than one match and keep only the match with the shortest Mahalanobis distance. The remaining de-matched treatment users were put back on queue to be re-matched. We also then remove those control users who already have a match from the control dataset such that the treatment user will have to find its next best match. This is repeated iteratively until conversion, when no control user has two matches.

Once the algorithm stops we have obtained a set of control users that equals the treatment users in number and is as similar to the pre-treatment treatment users as possible. We center treatment user data such that day 0 is the day each user became active in the hateful subreddit (i.e., when the treatment began). For control users their day 0 is the same as that assigned to their matched treatment user. This means both users in a given \{treatment user, control user\} pair have the same day 0, but the centering date differs between pairs.

\subsection{Interrupted Time Series}
ITS is a subset of Regression Discontinuity Design models which emphasizes specifically the modelling of effects over a continuous time period. It is well established and regarded for its usage in causal modeling in various fields. \cite{cattaneo_idrobo_titiunik_2020,Lee2010RDD,Ham2022}. 

Throughout this analysis, for the treatment user data we consider only the posts outside the studied subreddit, as we seek to understand the effects that joining one community has in the users' behavior outside of that community, and to control for in-group specific behavior. Additionally, the hate speech lexicons are overfit to each studied subreddit, which could affect results if the target subreddit was not removed from the treatment users data. The treatment subreddit is by definition not present in the data for control users.

ITS requires defining how many days around the treatment date (before and after) will be considered by the model. This analysis period is known as the bandwidth and current best-practices involve the use of cross-validation to find the optimal bandwidth \cite{Ludwig2005,Imbens2008}. Since we must be careful not to overfit the bandwidth choice to the treatment being analyzed, this procedure is done only on data prior to the treatment date (day 0). The cross-validation procedure consists of selecting several bandwidths to try (e.g. 30, 50, or 100 days), defining how many cross-validation rounds to run (e.g. 10, 25, 100), and an evaluation metric suitable for leave-one-out samples, with RMSE being the recommended choice and the one used in this analysis \cite{Jacob2012}. Then the cross-validation process works as follows: For a bandwidth of 50 days, with 10 cross-validation rounds, we would then fit a linear regression to days -51 to -2 and use the model to predict the \textit{proportion of hate speech} on day -1, then fit a regression on days -52 to -3 and predict on day -2, repeated up to predicting on day -10 \cite{Baicker2019}. From that set of predictions and known truth values we can then calculate the cross-validation RMSE for a given bandwidth. After iterating though all bandwidths, we select the one with the smallest RMSE. Under this procedure, testing bandwidths in the range [30,365] in increments of 5 days, and using 100 cross-validation rounds, we find the optimal bandwidths for each of the studied subreddits to be: r/GreatApes = 85 days; r/CoonTown = 30 days; r/fatpeoplehate = 55 days; r/Incels = 310 days. Bandwidths below 30 days were not studied due to their lower statistical power.

\textbf{Model Construction.} We fit a single Ordinary Least Squares (OLS) regression model to the entire dataset while using dummy variables to indicate which samples belong to the treatment and control groups, and which samples belong to the pre- and post-treatment groups \cite{Jacob2012}. In ITS lingo the treatment group is referred to as \textit{exposed}, such that \textit{exposed} = 0 means control group and \textit{exposed} = 1 means treatment group. The break-point in the time dimension when users begin to be treated is represented by the variable \textit{interrupted}, such that \textit{interrupted} = 0 means pre-treatment and \textit{interrupted} = 1 means post-treatment. In our case all samples up to but not including day 0 are pre-treatment samples (interrupted = 0), and all samples from day 0 (inclusive) onward are post-treatment samples (interrupted = 1). The OLS regression also includes the continuous \textit{time} variable, as well as all possible interaction terms between the three aforementioned variables (\textit{exposed}, \textit{interrupted}, and \textit{time}). The resulting set of coefficients is shown in Table \ref{tab:Coefficients}. 
%From the fitted coefficients, we can obtain a measure of the effect that the treatment has in the users, as well as how the user's hate speech is affected by the passing of time. 
Wald's F-test shows all fitted models are statistically significant (p-value $<10^{-20}$). 
%Since an ITS model is meant to be interpreted as a whole, the most appropriate measure is the Wald Test's F-Statistic, which gives both a score and an associated p-value that allows for the measurement of the goodness of fit for the entire model \cite{Baicker2019}. In the table \ref{tab:F-Statistic} we present such data for the four analyzed subreddits. All p-values are much lower than 0.05 ($<10^{-20}$), indicating that all ITS models have statistical significance.
%\begin{table}[h]
%\centering
%\begin{tabular}{l c} 
% Subreddit & F-Statistic \\ 
% \midrule
% r/GreatApes & 33.38  \\ 
% r/CoonTown & 40.99 \\ 
% r/fatpeoplehate & 26.31\\ 
% r/Incels & 61.40 \\ 
% \bottomrule
%\end{tabular}
%\caption{F-Statistic for the ITS model of each subreddit studied. P-values are $<10^{-20}$.}
%\label{tab:F-Statistic}
%\end{table}

\begin{table}[h]
\footnotesize
\centering
\begin{tabular}{ ll } 
 Parameter & Meaning \\ 
\midrule
 $const$                & Pre-treatment baseline \\ 
 $time$                 & Pre-treatment trend  \\ 
 $expos$                & Incremental baseline for the treatment group \\ 
 $inter$                & Incremental baseline after treatment\\
 $time \times expos$    & Incremental trend for the treatment group \\ 
 $time \times inter$    & Incremental trend after treatment \\ 
 $expos \times inter$   & Incremental baseline on treatment group \\
                        & after treatment \\ 
 $time \times expos \times inter$ & Incremental trend for the treatment group \\
                                    & after treatment \\
\bottomrule
\end{tabular}
\caption{Interpretation of coefficients obtained from the ITS model}
\label{tab:Coefficients}
\end{table}

 %The hate speech increase is defined as \textbf{interrupted + (exposed \times interrupted)}, where \textit{interrupted} measures any instantaneous changes to the baseline hate speech for all of Reddit (as represented by the control group), yet this value is expected to be zero, which is indeed the case as reported by the OLS regression parameters, and as can be seen on the ITS plots. Hence, most important among all these coefficients is \textbf{exposed \times interrupted} as it measures the amount of hate speech increase associated with the treatment users after they receive treatment, which is to say it captures the effect that joining the hateful subreddit has in a person's hate speech. 

\textbf{Model Interpretation.} The coefficients in Table \ref{tab:Coefficients} can be used to measure the relative increase in hate speech, defined as the difference between pre-treatment and post-treatment hate speech divided by the pre-treatment hate speech. For treatment users, the pre/post gap can be measured by \textit{$inter + expos \times inter$}, whereas the pre-treatment hate speech is defined as \textit{$const + expos$}. This relative measure looks at the instantaneous effect of joining the treatment subreddit, that is, it leverages the ITS's best estimate for the immediate increase that happens as a result of joining, while being unaffected by time-related trends in the data.

\textbf{Sensitivity Analysis.} We perform sensitivity analysis by fitting several OLS regression models using all bandwidths in the [30,365] range and then observing the coefficients for each of the ITS features as well as the corresponding p-values. This helps ensure that the results being considered are not outliers resultant from a very specific bandwidth size, and shows that all measured coefficients converge to consistent values regardless of bandwidth size. Sensitivity analysis confirms one challenge of ITS which was already to be expected: When using a small bandwidth the model is built with few samples, resulting in large variance and thus high p-values on the t-tests of individual coefficients. Then, as the bandwidth size increases the confidence intervals and p-values shrink.

\section{Results}

% To answer our research question ''\textbf{Does becoming active in a hateful subreddit increase a person’s usage of hate speech outside of that subreddit?}'' we employ the Interrupted Time Series (ITS) design (see Methods). The results obtained by the ITS design confirm our initial hypothesis that users increase the usage of hate speech in other subreddits after joining the hateful subreddit. 

\begin{figure}[h]
    \centering
    \includegraphics[width=0.9\linewidth]{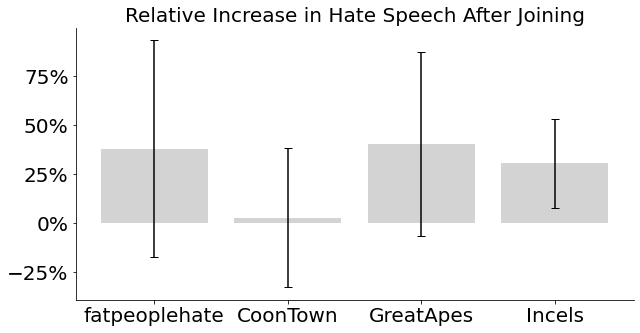}
    \caption{Relative increase in the rates of hate speech immediately after users join the hateful subreddit, as obtained from ITS model}
    \label{fig:subreddit_Comparison_All}
\end{figure}

The initial analyses show that the users who joined hateful subreddits will increase their usage of hateful speech outside of that subreddit by an average of $\approx 30\%$ immediately after joining. This estimate was made by using the individual coefficients obtained from the ITS model, and the pre/post gap measurement explained in the Model Analysis section under Methods. 
In more detail, users who joined r/CoonTown show the smallest increase in hate speech outside of r/CoonTown with just $3\%$ average increase, while the users who joined other three subreddits display substantially larger increase in hate speech usage with $38\%$, $40\%$, and $30\%$ for r/fatpeoplehate, r/GreatApes and r/Incels respectively.
The relative increase in hate speech immediately after users join the hateful subreddit is illustrated in Figure~\ref{fig:subreddit_Comparison_All}. Values are measured as the percentage increase of the users' hate speech immediately post-join compared to their immediately pre-join hate speech levels.

%how higher is treatment users' hate speech immediately post-join as a percentage of their immediately pre-join hate speech levels. 

%From the four subreddits studied, r/CoonTown sees a much smaller increase in hate speech, while the three other subreddits display an approximate 30\% increase in out-group hate speech as an immediate result of users joining the community.

\begin{figure}[h]
  \centering
  \includegraphics[width=1\linewidth]{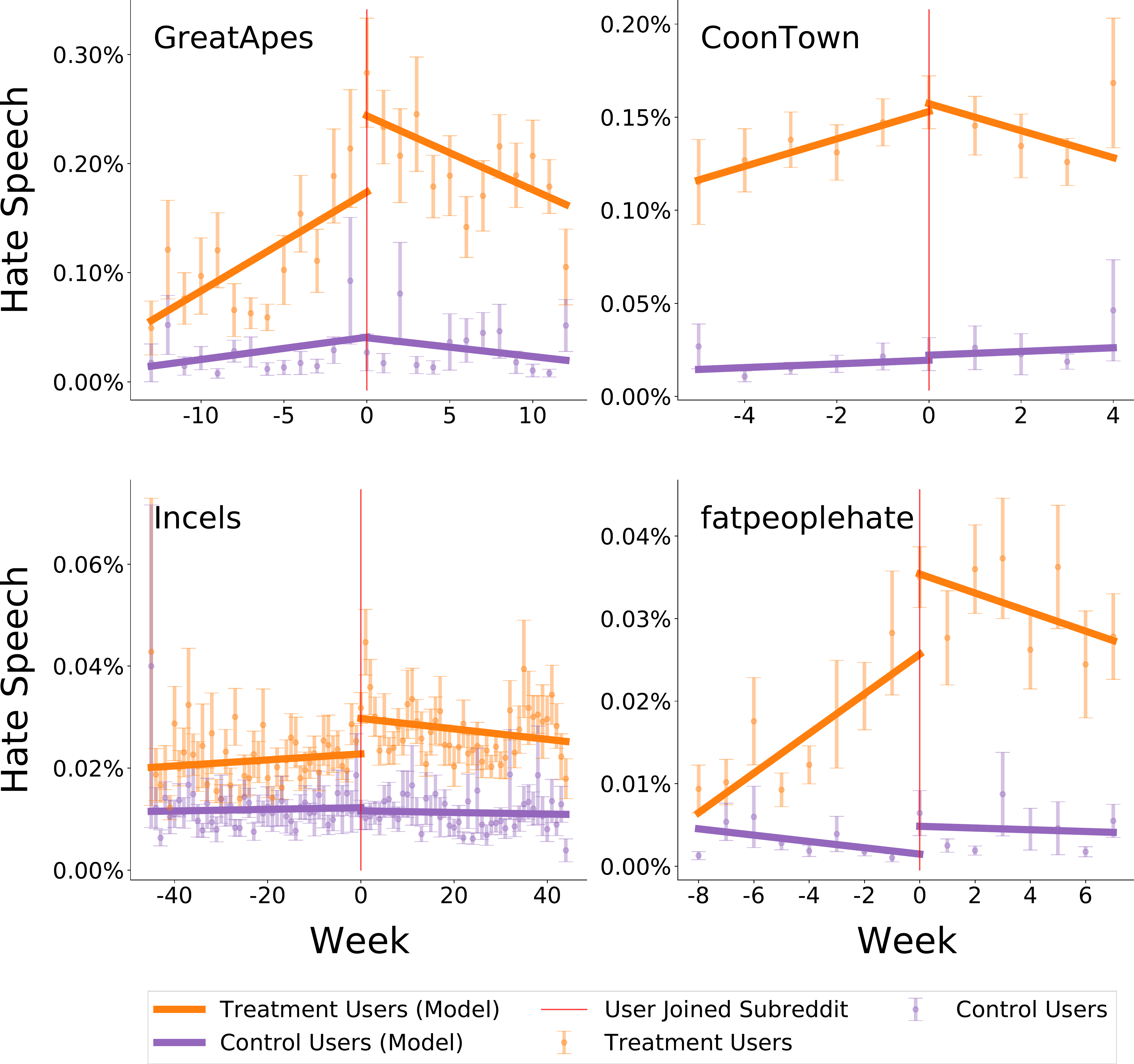}  
\caption{Interrupted Time Series plots for the studied subreddits.}
\label{fig:ITS}
\end{figure}

In addition to the immediate increase measured, there are further insights to be observed from plotting the results of the ITS model over time, as shown in Figure \ref{fig:ITS}. For each studied subreddit, combining the dummy variables (see Methods) through all possible interaction terms allows a single ITS model to generate four regression best-fit estimates of hate speech, both for treatment and control users, on both pre- and post-treatment periods. It is not surprising that the users in the treatment group (those who joined hateful subreddits) already had a relatively high rate of hate speech before joining the hateful subreddits compared to the control users. Still, the treatment groups show significant increase in rates of hate speech after joining the hateful subreddits, while the control users show no change. Such difference in trends suggests that joining the hateful subreddit has a major effect on the rate of hate speech expressed by the users who joined.

%Although there is an observed pre-treatment gap in hate speech between treatment and control users, that gap increases after users join the hateful subreddit. The absolute increase and the rate of hate speech over time for treatment and control groups is shown in Figure \ref{fig:ITS}.

From the plots we observe that, with varying intensities, treatment groups were already ramping up their hate speech before they joined the hateful subreddit. ITS design still allows us to conclude that the joining of that hate subreddit increases the user's hate speech immediately, but this trend nevertheless raises questions regarding what drives users to become members of hateful communities at the first place. From the ITS model, we observe a post-join downwards trend, indicating that the immediate post-join hate speech spike wanes over time, although at the end of the analysis period all subreddits were still above the hate speech values observed at the beginning of the analysis period. Fixing the hate speech baseline to the start of the analysis period, the end-of-period hate speech values saw increases of 189\%, 10\%, 25\% and 320\% for r/GreatApes, r/CoonTown, r/Incels, and r/fatpeoplehate respectively.

To ensure the ITS model shows robust results regardless of the time bandwidth, we additionally performed a Sensitivity Analysis (see Methods). Here, we vary the bandwidth and observe the change of various model coefficients. Across all bandwidths we observe a high p-value ($>$0.05) in the three parameters that measure external effects occurring on all users: $time$, $inter$, and $time \times inter$. Those high p-values suggest that the we cannot reject the null hypothesis that the true values of aforementioned variables are zero. Since the three coefficients obtained by ITS model are statistically indistinguishable from zero, we assume that we do not have evidence for any significant global (external) effects acting on all users, as those effects would be captured by the three aforementioned variables. Hence we conclude the control users express no change in their hate speech over time. This implies all measured effects are resultants of treatment users joining the hateful subreddit. The action of  joining, denoted as the $expos$ variable and its interaction terms, is the only difference between treatments and controls. In summary, thanks to the sensitivity analysis we show that the observed results are replicable across bandwidths, and not merely a by-product of an overfit optimal bandwidth.

We can observe a downwards trend in the treatment group post-joining. Given Reddit's active role in moderation, this downwards trend can, at least partially, result from Reddit's banning of individual users, which likely targets the most egregious offenders - those that most elevate the group's hate speech. Considering this common practice of banning users who express uncivil behavior, the average hate speech for the group is expected to decrease over time as only samples for the non-banned less-hateful members are still available.

To test this hypothesis we analyze the link between hate speech level and account lifespan -- the time an account is active on Reddit. We measure the average post-join hate speech for users who quit Reddit in less than one year after joining the studied subreddit and compare them to the users who stay longer. We observe that accounts with a longer lifespan have a lower average hate speech, as shown in Table \ref{tab:User-Lifespan}. 

\begin{table}[h]
\centering
\begin{tabular}{ lcc } 
  Subreddit & Up to 365 days & Over 365 days \\ 
 \midrule
 r/GreatApes & 0.421\% & 0.140\% \\ 
 r/CoonTown & 0.465\% & 0.101\% \\ 
 r/fatpeoplehate & 0.130\% & 0.001\% \\ 
 r/Incels & 0.083\% & 0.029\% \\ 
 \bottomrule
\end{tabular}
\caption{Average post-join hate speech for users with short and long lifespans on Reddit, as measured by the date of their last post in all of Reddit in relation to their becoming active on the hateful subreddit.}
\label{tab:User-Lifespan}
\end{table}

\begin{figure}[h]
  \centering
  \includegraphics[width=1\linewidth]{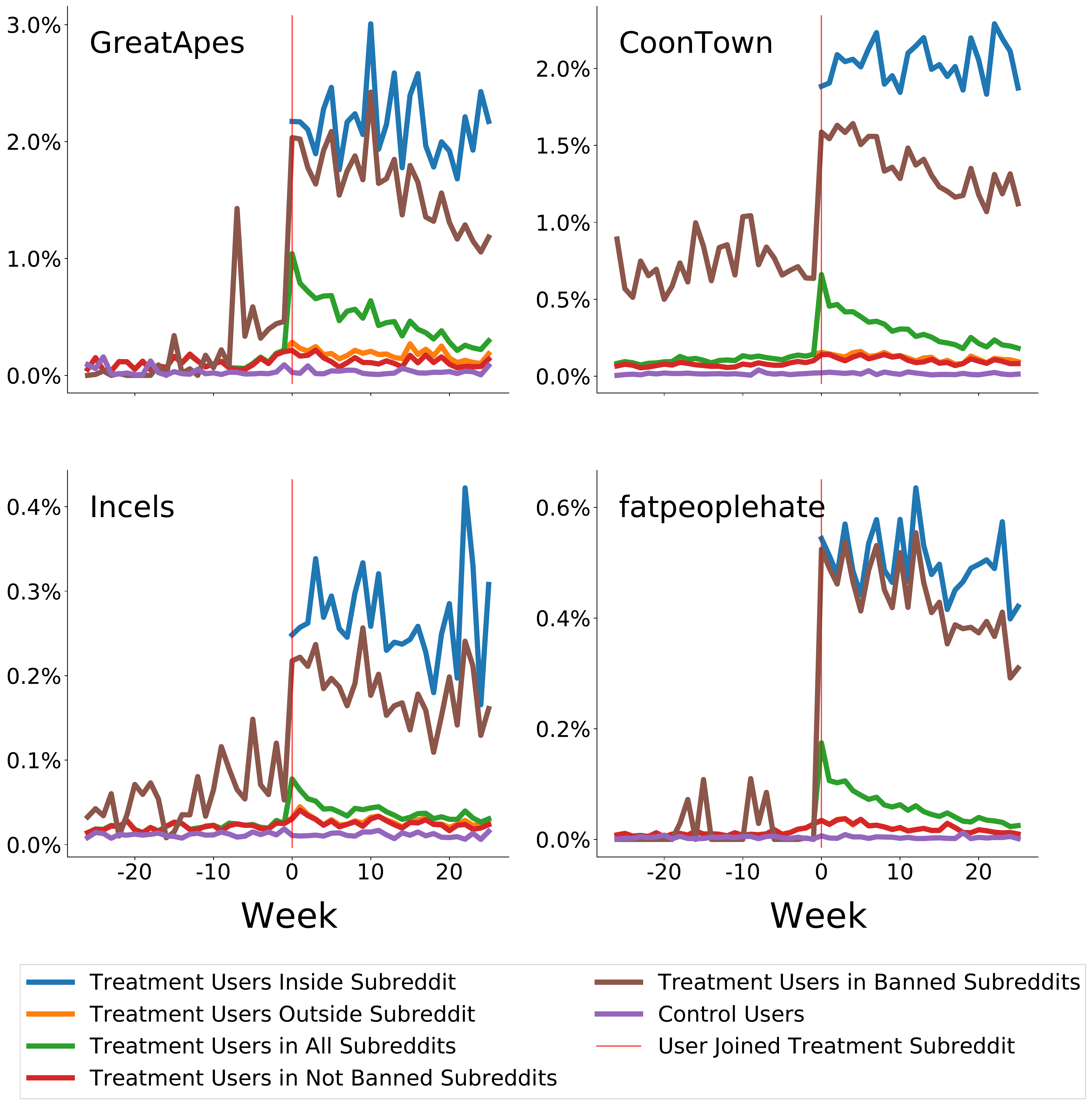}  
\caption{Temporal characteristics of expressed hate speech for treatment and control users in various scenarios.}
\label{fig:Timeseries}
\end{figure}

Complementarily, we leverage our data to explore hate speech trends over time for an expanded set of contexts in which treatment users can be found. We analyze the hate speech in the following scenarios: \textit{in all subreddits}, \textit{inside the treatment subreddit}, \textit{outside the treatment subreddit}, \textit{inside banned subreddits} and \textit{inside non-banned subreddits}. The plots illustrating the rate of hate speech of treatment users over time are shown in Figure \ref{fig:Timeseries}.

While effect sizes vary based on each subreddit and its unique lexicon, we generally observe that treatment users already had higher hate speech levels in the ``banned subreddits`` category, and that after joining the studied hateful subreddit, their hate speech displays a visually noticeable increase in the banned subreddits category. It also increases in the ``all`` category, which is to be expected since this includes the hateful subreddit itself.
Hate speech trends in the non-banned set closely tracks that of the outside set, which is to be expected as those are highly overlapping. Yet it is noteworthy that the time-series of both sets are still visually above that of the control set.

\begin{figure}[h]
  \centering
  \includegraphics[width=1\linewidth]{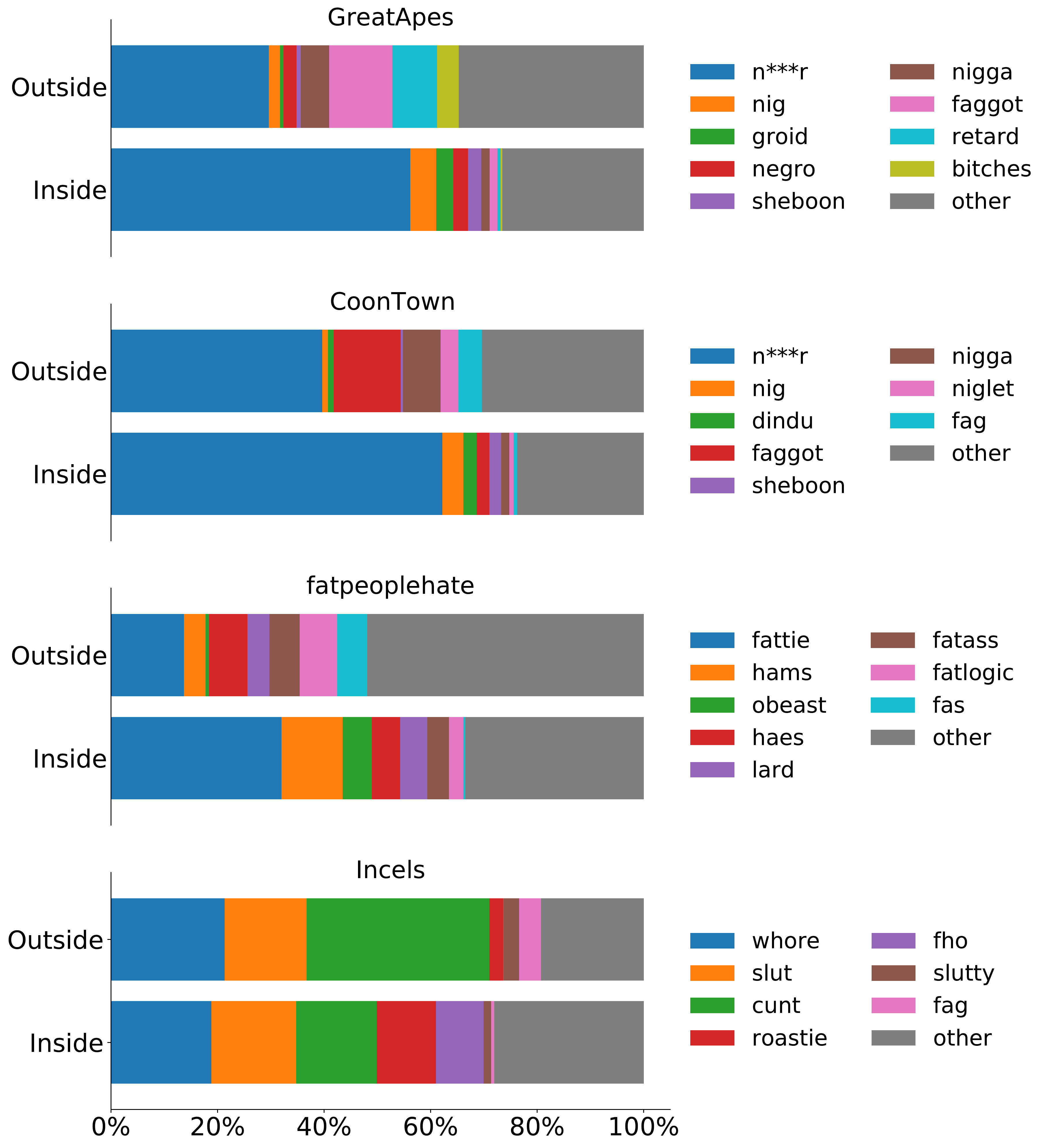}  
\caption{Distributions of the most frequent hate words used inside and outside of each observed subreddit}
\label{fig:WF}
\end{figure}

All analysis were done based on the custom lexicons developed during this study. Figure \ref{fig:WF} illustrates the lexicons' most frequently written hate words in the in-group and out-group settings, considering posts after the users became active members of the hateful subreddit. We compare both distributions via a Spearman Rank Correlation test, which measures the monotonicity of the relationship between two datasets, here represented as the in-group and out-group word distributions. For all four subreddits studied the test obtained a p-value below 0.05, meaning we can reject the hypothesis that both word distributions are generated by the same underlying system. This confirms there is a difference in the distributions with which hate words are chosen when communicating to the in-group versus out-group.

When looking at racism lexicons, the usage of the n-word nearly doubles when members are communicating with the in-group. Similarly significant increases happen with other highly offensive words such as ``obeast`` in the fat-shaming context. We also observe a reduction in group slang such as ``sheboon`` and ``roastie`` when speaking outside the hateful community. This effect happens across all categories of hate speech studied. Conversely, the out-group data shows an increase in derogatory-yet-not-as-societally-shunned terms such as ``fatlogic`` and ``fag``. In summary, discoursing in the out-group displays a relative toning-down of the most offensive words and of insider slangs, counter-weighted by a relative increase in milder hate words.

% \begin{figure*}[h]
% \begin{subfigure}{.5\textwidth}
%   \centering
%   % include first image
%   \includegraphics[width=1\linewidth]{figures/Word_Frequencies_Pie_Chart_GreatApes.png}  
%   %\caption{Put your sub-caption here}
%   \label{fig:WF-GreatApes}
% \end{subfigure}
% \begin{subfigure}{.5\textwidth}
%   \centering
%   % include second image
%   \includegraphics[width=1\linewidth]{figures/Word_Frequencies_Pie_Chart_CoonTown.png}
%   %\caption{Put your sub-caption here}
%   \label{fig:WF-CoonTown}
% \end{subfigure}
% \begin{subfigure}{.5\textwidth}
%   \centering
%   % include third image
%   \includegraphics[width=1\linewidth]{figures/Word_Frequencies_Pie_Chart_fatpeoplehate.png}
%   %\caption{Put your sub-caption here}
%   \label{fig:WF-fatpeoplehate}
% \end{subfigure}
% \begin{subfigure}{.5\textwidth}
%   \centering
%   % include fourth image
%   \includegraphics[width=1\linewidth]{figures/Word_Frequencies_Pie_Chart_Incels.png}  
%   %\caption{Put your sub-caption here}
%   \label{fig:WF-Incels}
% \end{subfigure}
% \caption{Most frequent hate words in each subreddit}
% \label{fig:WF}
% \end{figure*}

\section{Discussion}

Through a robust ITS analysis we were able to identify a significant increase in hate speech resulting from joining subreddits known for their toxicity. This increase was observed outside of the community which incited it, thus implying a spread of hate arising from joining hateful communities.

When analyzing the ITS models in Figure \ref{fig:ITS}, we noted that there is an upwards trend in hate speech that predates the moment when treatment users become active members of the hate community. It seems probable that that users were already becoming attracted to that specific thematic, perhaps already being passive members of the studied community, known in Reddit lingo as ''lurkers'', or users who read a subreddit's forum but do not engage in discussion. Due to dataset limitations, we measure the treatment start as the date in which a user made a first post in the studied subreddit. There is a looming possibility that treatment users were already lurkers before becoming active, and thus their external behavior had already began reflecting their increasing extremism.

Additionally, we observe a downward trends in the treatment group post-joining. We find an evidence for a link between the decreased hate speech and the accounts' lifespan. The lifespan is affected by either a voluntary abandonment of accounts created for usage on the hateful community, or banning by Reddit, although our data does not permit to differentiate amongst those possibilities. Besides the effects of Reddit's banning polices, that were briefly explored in this study, there could be other reasons for such phenomenon. A commonplace explanation could be a simple regression to the mean -- a phenomenon in which those performing extremely well or extremely poorly (outliers) tend to move closer to the mean over time. Further research is needed to confirm this hypothesis. 

When looking at the percentage of hate speech produced by the members of the studied hateful subreddits in various contexts, it is observable that those users' hate speech also increases in other subreddits that were also eventually banned at a later date as illustrated in Figure \ref{fig:Timeseries}. We can see an expected connection between a subreddit adopting hateful words, and a probability of it being banned. Moreover, since a subreddit cannot adopt hate words after it is already banned, it must be that the adoption of hate words precedes the ban, although this analysis cannot necessarily prove that the adoption causes the ban.

%\subsection{In-Group Versus Out-Group Hate Vocabulary}
Regarding hate speech vocabulary, in Figure \ref{fig:WF} we observe clear differences in hate word choices when subreddit members express themselves inside versus outside the community. Those differences were confirmed by a Spearman Rank Correlation test. We observe that users choose more ``insider`` hate words within their community, but less frequently bring these slang to the outside world, which is in line with some previous findings \cite{trujillo2021echo}. Such behavior might be one way of evoking group membership, by displaying knowledge of the group's unique vocabulary, which has been shown to entice greater responsiveness from the community \cite{tran2016characterizing}. The converse is also true, where users who employ more obscure in-group language outside of their group might simply find low engagement in other forums, either due to the out-group's disinterest in a foreign cause, or simply due to others inability to comprehend messaging laden with group-specific language. Both scenarios would discourage the usage of group-specific language in favor of a more commonplace vocabulary and style when communicating outside of the users' community.

From a qualitative perspective, in both the racist and the fat-shaming communities there appears to be a toning-down on the relative usage of the most egregiously offensive words, such as the n-word and ``obeast``, when communicating out-of-community. A possible reason to explain this is that even hate group members are still somewhat affected by society's pressure towards civility. A secondary factor might once again be moderation, as those who chose the most offensive words in the out-group would be at higher likelihood of having their posting ability blocked, thus halting their spread of such vocabulary. %, and leaving the majority of hate speech experienced in the platform limited to that which is not sufficiently shunned down onto to get its propagator blocked from posting other communities.

From observing four different subreddits, covering three categories of hate speech, we have shown a causal relationship between a user becoming active in the community and the user's hate speech increasing immediately after. Such negative externality that arises from the growth of hateful communities makes a case for stronger moderation of such communities, an area where quarantining subreddits shows some promise. In general these findings corroborate with previous research \cite{Chandrasekharan2017,trujillo2021echo} that suggests positive hate reducing effects can be obtained by restraining such hateful communities.

\section{Conclusion}
In this paper we presented a causal inference analysis of the impact that joining a hateful subreddit has on a user's ensuing hate speech in other subreddits. We looked at Reddit, one of the major social media platforms, and discover that the hate speech increases as an effect of joining hateful subreddits. We find the positive effects in three different categories of hate speech: racism, misogyny and fat-shaming. 

%While much has been researched about social media, most findings have been in the realm of correlation. In this regard 
This study displays novelty for using causal modeling to understand the influence of social media in hate speech, especially by leveraging ITS design to identify the effects. Additionally, we perform the sensitivity analyses by exploring the optimal time bandwidths and assessing the modulating effects of different time bandwidths on calculated effect sizes.

% \subsection{Limitations}
However, potential limitations warrant consideration. Our findings indicate that despite the matching technique employed, there were still pre-existing differences between the treatment and control groups, which stem at least in part from the need of matching without using the target variable, as commonly expected from such approaches \cite{Ham2022}. This causes the treatment group to have a higher baseline hate speech level, measured as \textit{const} + \textit{expos}. Nevertheless the ITS model utilized is robust to different baselines by including a control group precisely to allow measurement of such extraneous effects. The approach to lexicon construction or the choice of lexicon usage can also have an impact in the analyses as hate speech is not binary, but rather covers an entire spectrum, from positively talking about the out-group, passing through mild-offenses and ending in a toxic discourse. Certain outlets might be more conducive to different extents of hate speech and thus differences could be found when lexicons differ.

We were also limited in our data availability, as it is only possible to measure a user's active actions, namely posting and commenting. This constraint makes analysing the effects of passive content consumption challenging. Similarly, we only study four subreddits, and the additional analyses is needed to assess the generality of our conclusions in Reddit and other social media platforms.

Given the inability to obtain data regarding when a user formally chooses to become member of a subreddit, nor data regarding which posts were being read by a user, the current approach could not investigate if those relate to the observed pre-join upwards trend, thus leaving opportunity for future work. The mirror case is also true, exploration of what is driving the treatment users speech downwards after joining the hateful subreddit could unearth additional information to be leveraged in combating hate speech.

\section*{Acknowledgements}
Funding for this work is provided through the USC-ISI Exploratory Research Award and through DARPA (Award \# HR0011260595 and \# HR001121C0169)

\section*{Conflicts of Interest}
The authors declare no conflicts of interest.
    
\bibliographystyle{plain}
\bibliography{references.bib}

\end{document}